\documentclass[12pt]{article}
 \usepackage[cp866]{inputenc}
  \usepackage[english]{babel}

\usepackage{latexsym}
\usepackage{amsmath}
\usepackage{amssymb}
\usepackage{graphicx}
\usepackage{indentfirst}

\newcommand{\be}{\begin{equation}}
\newcommand{\ee}{\end{equation}}

\newcommand{\bea}{\begin{eqnarray}}
\newcommand{\eea}{\end{eqnarray}}
\newcommand{\nn}{\nonumber}

\newcommand{\bk}{{\bf k}}
\newcommand{\bv}{{\bf v}}
\newcommand{\bp}{{\bf p}}
\newcommand{\br}{{\bf r}}
\newcommand{\bq}{{\bf q}}
\newcommand{\bQ}{{\bf Q}}
\newcommand{\bb}{{\bf b}}

\newcommand{\la}{\langle}
\newcommand{\ra}{\rangle}

\usepackage[dvips]{color}

\newcommand{\dst}{\displaystyle}
\newcommand{\fr}[2]{\frac{{\dst #1}}{{\dst #2}}}

\date{July 29, 2015}

\title{Born approximation for scattering \\ of wave packets on
atoms\\
 {\it I. Theoretical background for scattering \\of a wave packet on a potential field}}

\author{D.V.~Karlovets$^{1}$, G.L.~Kotkin$^{2,3}$, V.G.~Serbo$^{2,3}$, 
\\
      {{\small $^1$ Tomsk Polytechnic University, Lenina 30, 634050 Tomsk, Russia}}\\
		{\small $^2$ Novosibirsk State University, RUS-630090, Novosibirsk, Russia}\\
{\small $^3$ Sobolev Institute of Mathematics, RUS-630090, Novosibirsk, Russia}\\
    }

\begin{document}

\maketitle
\begin{abstract}

Laser photons carrying non-zero orbital angular momentum
are known and exploited during the last twenty years. Recently it has
been demonstrated experimentally that such (twisted) electrons can
be produced and even focused to a subnanometer scale. Thus, twisted
electrons emerge as a new tool in atomic physics. The state of
a twisted electron can be considered as a specific wave packet of
plane waves. In the present paper-I we consider  elastic
scattering of the wave packets of fast non-relativistic particles on a
potential field. We obtain simple and convenient formulae for
a number of events in such a scattering. The equations  derived represent,
in fact, generalization of the well-known Born approximation
for the case when finite sizes and inhomogeneity of the initial
packet should be taken into account. To illustrate the obtained
results, we consider two simple models corresponding to scattering
of a Gaussian wave packet on the Gaussian potential and on the
hydrogen atom. The scattering of twisted electrons on atoms will
be considered in the next paper-II.

\end{abstract}

\section{Introduction}

Let us consider a scattering of a particle beam on a potential
field. The real beam has  finite sizes and nonuniform density,
but usually in the standard theoretical description of the scattering
process, the beam is replaced by a plane wave. Such an approach is
valid for a number of problems in which distances, essential for
calculation of the corresponding cross sections, are
considerably smaller than the typical sizes of beam inhomogeneity.
However, there are important exclusions.

Thus, at $e^+e^-$ and $ep$ colliders several processes were experimentally
investigated and then calculated theoretically in which macroscopically large impact parameters gave an essential contribution to the cross section. These impact
parameters may be much larger then the transverse sizes of the
colliding bunches. In that case, the standard calculations have to
be essentially modified. It is the so called beam-size or MD-effect,
discovered at the MD-detector (the VEPP-4 collider in Novosibirsk)
--- see review~\cite{KSS-1992}. This effect is important because it
leads to the essential reduction of beam particle losses at modern
colliders~\cite{LEP-lum}. Another important example here is the so-called pre-wave zone effect
in transition radiation and diffraction radiation of electron beams \cite{V}, which is of high importance for beam diagnostic techniques in modern accelerators \cite{DR}. We should note that these phenomena are directly related to the ultra-relativistic motion of particles.

In the present paper we discuss non-relativistic elastic
scattering of wave packets on the potential field in a  situation
when sizes of the initial packet may be comparable to the typical
radius of the field action. This problem becomes especially topical due
to recent experiments with the twisted electrons. These are states of
the beam whose electrons have a defined value $\hbar m$ of the
{\it orbital angular momentum (OAM) projection} on the beam
propagation axis. Laser beams carrying OAM are well known in
optics,~\cite{OAM} (for a review see \cite{OAMreview}). Numerous
applications of light with the OAM are described in the recent
book~\cite{TP}. History of twisted electrons is shorter. Indeed,
only a few years ago, following the suggestion made in
\cite{bliokh2007}, several groups reported successful
creation of the twisted electrons, first using the phase plates
\cite{twisted-electron} and then with the computer-generated holograms
\cite{twisted-electron2}. Such electrons carried the energy  up to 300 keV and the orbital quantum number up to $m= 200$ \cite{GrG15}.
These vortex beams can be manipulated and focused just as the
conventional electron beams, and recently remarkable focusing of a
twisted electron beam to a focal spot of less than $0.12$ nm in
diameter was achieved~\cite{Ant-2011}. It means that wave packets
in the form of twisted electrons emerge as a new tool in atomic
physics. One of the first atomic processes with twisted electrons --
the radiative capture of twisted electrons by bare ions with the
emission of a photon -- was recently studied in the theoretical
paper~\cite{MHSSF-2014}.

We would like to emphasize that while the use of the wave-packets instead of plane waves may be usually avoided in the standard approach (see e.g. Chapt. 4.5 in \cite{PS}), this seems to be no longer the case for scattering problems with the twisted electrons and photons. Indeed, the use of the pure Bessel (non-normalizable) states was shown to be inconsistent with the conservation law of the OAM in a $2\rightarrow 2$ scattering, and only the formalism of the well-normalized wave packets removes this difficulty \cite{IS}. 

The next important theoretical problem is to study the elastic scattering of twisted electrons on atoms and atomic structures. The first theoretical investigations in this field \cite{VaB13,BoP15,SIFSS-15} concentrated on some important features of such processes but do not address the generic issue: how to calculate the number of events depending on the limited sizes of the incident beam. To approach this goal, we derive a simple and convenient expression for the number of events, which generalizes the well-known Born approximation for the case when the initial beam
is a wave packet, but not a plane wave. New and important feature
of this problem is the fact that the observed number of events
depends on the impact parameter $\bb$ between the potential centre
and the packet axis. We also consider a case when the wave packet
is scattered on the randomly distributed potential centres. From the
experimental point of view, this case is the simplest one.
For such a set-up we obtain a simple and transparent expression for the cross section averaged over impact parameters.

To illustrate the general formulae obtained, we consider in the
present paper-I two simple models:

1) Scattering of the Gaussian wave packet on the Gaussian
potential;

2) Scattering of the Gaussian wave packet on the hydrogen atom.

The detailed analysis of the angular distribution of final
particles is given. Special attention is paid to the dependence of
this distribution on the impact parameter of the potential centre
with respect to the packet axis. It should be noted that the models being discussed have an advantage of simplicity, that is why some new features in
scattering of the wave packets we study first with these models and only
after that we proceed with more realistic models of the twisted electrons' wave packets.

More interesting and more sophisticated scattering of the twisted
electrons on atoms will be considered in the next paper-II.

The structure of the paper-I is the following. In the next section
we remind the standard Born approximation including well-know
formulae for the scattering on the Gaussian potential and on the
hydrogen atom. In Sect. 3 we derive the basic formulae for
scattering of a wave packet on a potential field, in Sect. 4 we
specify the general approach for the case of the Gaussian wave
packet and illustrate this case by detailed consideration of two
above mentioned models. Some conclusion is given in Sect. 5.

For definiteness, we consider below the central fields $U(r)$ having in mind that the presented method can be applied for non-central fields $U(\br)$ as well.

To simplify formulae, we use units with the Plank constant $\hbar =1$.

\section{Standard Born approximation}

\subsection{Number of events in the standard Born approximation}

Let us consider the scattering of a packet of non-relativistic
particles (electrons, for definiteness) off a potential field
$U(r)$ whose centre is located at the coordinate origin. Let the
typical radius of this field's action be of the order of $a$. If
the initial and scattered electrons have momenta $\bp_i$ and
$\bp_f$, then the $S$-matrix element $\la f|S|i \ra$ for the
transition between plane waves $|i\rangle =|\bp_i\rangle$ and
$|f\rangle =|\bp_f\rangle$ is expressed via the scattering
amplitude $f(\varepsilon_i, \theta, \varphi)$ as follows:
 \be
\la f|S|i \ra=(2\pi)^2 i
\,\delta(\varepsilon_i-\varepsilon_f)\,\fr{f(\varepsilon_i,
\theta, \varphi)}{m_e}\,,
\;\;\varepsilon_i=\fr{\bp_i^2}{2m_e}\,,\;\;
\varepsilon_f=\fr{\bp_f^2}{2m_e}\,,
 \ee
where $\theta$ and $\varphi$ are the polar and azimuthal scattering angles and $m_e$ is the electron mass. The standard
differential cross section of the process equals
 \be
\fr{d\sigma_{\rm st}}{d\Omega}=|f(\varepsilon_i, \theta, \varphi)|^2\,,
 \label{sts}
 \ee
where $d\Omega$ is the solid angle element. The scattering
amplitude in the first Born approximation is related to the
Fourier transform of the potential field  (see, for example, \S 126
in the text-book \cite{LL-KM}):
 \be
f(\bq)=-\fr{m_e}{2\pi}\,\int U(r)\,e^{-i\bq\,\br}\, d^3r\,,\;\;
\bq=\bp_f-\bp_i\,.
 \label{Born}
 \ee

In the standard approach there is an implicit assumption that the
particle bunch is wide and long and almost uniform on distances of
the order of $a$, i.\,e. the bunch density in the region of
active forces is $n(\br, t)\approx n({\bf 0}, t)$, and velocities
of all electrons are almost equal to each other and directed along
the axis $z$, therefore,
 $$
\bp_i=m_e\bv_i=(0,\,0,\,p_i)\,.
 $$
In such a case the number of particles  $d\nu$ scattered over the time
$dt$ is determined by a product of the differential cross section
$d\sigma_{\rm st}$ and the current of particles near the coordinate
origin for a given time $v_i n({\bf 0},t) dt$, while the total
number of the scattered particles for the whole time
reads\footnote{For colliding beams, the quantity analogous to $L$
is called {\it luminosity}.}
 \be
\fr{d\nu_{\rm st}}{d\Omega} = L \,\fr{d\sigma_{\rm st}}{d\Omega}=
L\,|f(\bq)|^2,\;\; L=\int v_i n({\bf 0},t)
dt\,.
 \label{stannu}
 \ee

Let the initial state be the plane wave in a large volume ${\cal
V=\pi R}^2 l_z$, where ${\cal R}$ is the radius and  $l_z$ is the
longitudinal length of the bunch. The bunch density during the
large time $\Delta t=l_z/v_i$ is almost constant and equals
$n(\br, t)= {N_e}/{\cal V}$, therefore,
 \be
L=\fr{N_e}{\cal V}\, l_z=\fr{N_e}{\pi {\cal R}^2}.
 \label{Lplane}
 \ee
Usually, the change of the transverse beam sizes during
the scattering can be neglected and the bunch density depends on the time as
$n({\br},\,t)=n({\br}_\perp,\,z-v_it)$. If we define {\it the transverse density}
 \be
n_{\rm tr}(\br_\perp)=\int n(\br, t) \,dz\,,
 \label{tr}
 \ee
then quantity $L$ coincides with the transverse density at the coordinate origin,
 \be
L=n_{\rm tr}(\br_\perp={\bf 0})\,.
 \label{lum}
 \ee

Let us remind two important examples (see, for example, problems
to \S 126 in the text-book \cite{LL-KM}).

\subsection{Gaussian potential}

The Gaussian potential has the form
 \be
U(r)=V\,e^{-r^2/(2a)^2}\,.
 \label{Gauss}
 \ee
If electrons are fast ($p_ia\gg 1$) and the condition $V\ll
p_i/(m_ea)$ is satisfied, then the Born amplitude equals
 \be
f(\bq)=f_0\,e^{-(qa)^2}\,,\;\; f_0=-4\sqrt{\pi} m_e Va^3\,.
 \ee
The total cross section is determined by the small angle region
$\theta \lesssim 1/(p_ia)$ and reads
 \be
\sigma_{\rm st}=\fr{\pi f_0^2}{2a^2p_i^2}\,.
 \ee

\subsection{Hydrogen atom in the ground state}

The scattering of fast electrons on the hydrogen atom in the
ground state is directly related to the scattering on the potential
field of the form (see \cite{LL-KM}, problem 2 to \S 36)
 \be
U(r)= -\fr{e^2}{r}\left(1+\fr{r}{a}\right)\,e^{-2r/a}\,,
 \label{atom}
 \ee
where $e$ is the proton charge and $a = 1/(m_e e^2)$ is the Bohr radius. The
Born approximation for fast electrons  ($p_ia\gg 1$) is valid if
$m_e e^2\ll p_i$, so the scattering amplitude is equal to
\be
 f(\bq)= \fr{a}{2}\,
 \left[\fr{1}{1+(qa/2)^2}+\fr{1}{(1+(qa/2)^2)^2}\right] \,,
 \label{f-atom}
 \ee
and the total cross section is
 \be
\sigma_{\rm st}=\fr{7\pi}{3\,p_i^2}\,.
 \ee

\section{Scattering of a wave packet on a potential field}

\subsection{Basic formulae}

In this section we follow the approach developed in Sect. 4.1 from
the paper~\cite{KSS-1992}. The initial state of incoming electrons
is given by the wave packet of the form
 \be
\int |\bk \ra \Phi(\bk) \,\fr{d^3k}{(2\pi)^{3/2}}\,,
 \ee
where the packet's wave function in the momentum space
$\Phi(\bk)$ is normalized by the condition
 \be
\int |\,\Phi(\bk)|^2\, d^3k=1\,.
 \ee
As a consequence, the probability amplitude for the transition
from this initial state to the final plane-wave state $|\bp_f \ra$
is given by the convolution
 \be
A=\int \la\bp_f|S|\bk\ra \Phi(\bk)
\,\fr{d^3k}{(2\pi)^{3/2}}=\sqrt{2\pi}\, i \int
\delta(\varepsilon-\varepsilon_f)\, f(\bp_f-\bk)\, \Phi(\bk)
\,\fr{d^3k}{m_e}\,,\;\; \varepsilon=\fr{\bk^2}{2m_e}\,,
 \ee
while the number of scattered particles equals
 \be
d\nu=N_e A\,A^*\,\fr{d^3p_f}{(2\pi)^3}\,,
 \label{nu}
 \ee
where $N_e$ is the number of electrons in the initial packet and
 \be
A^*=-\sqrt{2\pi}\, i \int \delta(\varepsilon'-\varepsilon_f)\,
f^*(\bp_f-\bk')\, \Phi^*(\bk') \,\fr{d^3k'}{m_e}\,,\;\;
\varepsilon'=\fr{(\bk')^2}{2m_e}\,.
 \ee

Taking into account that $d^3p_f=p^2_fdp_fd\Omega$, one can
perform integration over the longitudinal momenta $k_z$ and $k'_z$
using the $\delta$-functions and obtain
 \be
\fr{d\nu}{d\Omega}=N_e\int f(\bp_f-\bk)\, \Phi(\bk)
f^*(\bp_f-\bk')\, \Phi^*(\bk')\,\fr{p_f^2}{\tilde k_z \tilde
k_z'}dp_f\, \,\fr{d^2k_\perp}{2\pi} \,\fr{d^2k'_\perp}{2\pi},
  \ee
where
 \be
\bk=\left(\bk_\perp,\tilde k_z=\sqrt{p_f^2-k_\perp^2}\,\right),
\;\;\bk'=\left(\bk'_\perp,\tilde
k'_z=\sqrt{p_f^2-(k'_\perp)^2}\,\right).
 \ee

On the next step we simplify this expression using several natural
assumptions. We assume that the wave packets considered have an axial
symmetry, therefore, their averaged transverse momentum is zero,
$\la \bk_\perp \ra={\bf 0}$, their averaged momentum is
 \be
\la \bk \ra=\bp_i=m_e\bv_i=(0,\,0,\,p_i),
 \label{pi}
 \ee
but the averaged absolute value of the transverse momentum is non-zero,
 \be
\la k_\perp\ra=\varkappa_0 = p_i \,\tan{\theta_k}.
 \ee
Here we introduce the angle $\theta_k$, which is important
parameter for the twisted state and usually is called {\it the
conical or opening angle}. The packet axis can be shifted in the transverse
$xy$ plane by a distance (by an impact parameter) $b$. Below we
often choose the $x$-axis just along this shift --- in this case
${\bf b}=(b,\,0,\,0)$, the azimuthal angle $\varphi$ coincides
with the angle between vectors $(\bp_f)_\perp$ and $\bb$, and the
azimuthal angle $\varphi_k$ coincides with the one between
vectors $\bk_\perp$ and $\bb$.

The packet's wave function in the momentum space can be
presented as a product of wave functions corresponding to
the transverse and longitudinal  motions:
 \be
\Phi(\bk)=\Phi_{\rm tr}(\bk_\perp)\, \Phi_{\rm long}(k_z)
 \ee
with the dispersions $\Delta k_x=\Delta k_y \sim 1/\sigma_\perp$,
$\Delta k_z\sim 1/\sigma_z$, where $\sigma_\perp$ and $\sigma_z$
are the transverse and longitudinal averaged sizes of the packet.
We assume further that these dispersions are small compared to
the longitudinal momentum:
 \be
\Delta k_x=\Delta k_y\sim 1/\sigma_\perp\ll p_i\,,\;\;\Delta k_z
\sim 1/\sigma_z \ll p_i\,.
 \label{ineq-p-1}
 \ee

From the experimental point of view, it is interesting to consider
a case when the packet's length $\sigma_z$ is larger then the
radius of the field action $a$, but still small enough to provide such
a situation that during the collision time $t_{\rm col}\sim
{\sigma_z}/{v_z}= {m_e\sigma_z}/{p_i}$ the wave packet does not
spread essentially in the transverse plane. It means that the
collision time has to be considerably smaller than the diffraction
time $t_{\rm dif}\sim {\sigma_\perp}/{v_\perp} =\,
{m_e\sigma_\perp}/{\varkappa_0}$. Therefore, below we assume that
 \be
a\ll \sigma_z \ll {\sigma_\perp} \fr{p_i}{\varkappa_0}\,.
 \label{ineq}
 \ee

For further integration over $p_f$ or
$\tilde{k}_z=\sqrt{p_f^2-k_\perp^2}$ we take into account the
following properties of functions under the integral. The
amplitude $f(\bp_f-\bk)$ is concentrated near the value
$\tilde{k}_z=({\bp_f})_z=p_f \cos{\theta}$ with the dispersion
$\sim 1/a$, while the function $\Phi_{\rm long}(\tilde{k}_z)$ is
concentrated near the value $\tilde{k}_z=p_i$ with the dispersion
$\sim 1/\sigma_z$, which is considerably smaller than $1/a$.
Finally, we take into account that the quantities $\tilde k_z$ and
$\tilde k_z'$ depend on $k_\perp$ and $k_\perp'$ but the
corresponding variations are small, for example,
 \be
|\delta\tilde k_z|= \left|\delta\sqrt{p_f^2-k_\perp^2}\, \right|\sim \fr{\varkappa_0 \delta k_\perp}{
p_i}\lesssim \fr{\varkappa_0}{\sigma_\perp \, p_i} \ll
\fr{1}{\sigma_z}\sim \Delta k_z
 \ee
due to Eq. \eqref{ineq}. Therefore, we can take the amplitudes
$f(\bp_f-\bk) f^*(\bp_f-\bk')$ out of the integral over $p_f$ in the
form $f(\bp_f-\bp_i-\bk_\perp) f^*(\bp_f-\bp_i-\bk'_\perp)$ with
$\bp_i$ given in Eq.~\eqref{pi}.

The rest integral over $p_f$ can be evaluated as follows
 \be
\int \Phi_{\rm long}(\tilde k_z)\Phi_{\rm long}^*(\tilde
k'_z)\fr{p_f^2}{\tilde k_z \tilde k_z'}dp_f=\int \left|\Phi_{\rm
long}(\tilde k_z)\right|^2 \fr{p_f}{\tilde k_z} d\tilde
k_z=\fr{1}{\cos{\theta_k}}\,.
 \ee
As a result, we obtain the basic expression
 \be
\fr{d\nu}{d\Omega}=\fr{N_e}{\cos{\theta_k}}\,|F(\bQ)|^2\,,\;\;
F(\bQ)=\int f(\bQ-\bk_\perp)\, \Phi_{\rm tr}
(\bk_\perp)\,\fr{d^2k_\perp}{2\pi}\,,
 \label{nuf}
 \ee
where $\bQ$ is given as
 \bea
\bQ&=&\bp_f-\bp_i=(\bQ_\perp,Q_z),\;\;\bQ_\perp=(\bp_f)_\perp= p_f
(\sin{\theta}\cos{\varphi}, \sin{\theta}\sin{\varphi},0),
 \\
Q_z&=&p_f\cos{\theta}-p_i,\;\;p_f= \sqrt{p_i^2+\varkappa_0^2}\,.
  \nn
 \eea

Let us stress that the formula~\eqref{nuf} is valid only under
the condition~\eqref{ineq} --- in other words, this approximation is
inapplicable for too small values of $\sigma_\perp$. Besides, we would like to note the following. To obtain this formula we have used a non-monochromatic initial packet. Moreover, in the final state we integrate over the plane waves with different energies which are detected in the solid angle $d\Omega$. Our final
result~\eqref{nuf} includes the Born scattering amplitude
$f(\bp_f-\bp_i-\bk_\perp)$ in which the initial plane wave has the
momentum $\la\bk\ra+\bk_\perp=\bp_i+\bk_\perp$, while the final
plane wave has the momentum $\bp_f$ with $|\bp_f|=\sqrt{p_i^2+\la
k_\perp\ra^2}= \sqrt{p_i^2+\varkappa_0^2}$.

Below another representation for the integral $F(\bQ)$ will be useful. It can be obtained if we substitute the evident form of the scattering amplitude from~\eqref{Born} into Eq.~\eqref{nuf}:
 \be
F(\bQ)=-\fr{m_e}{2\pi}\int  U(r) \,\Psi_{\rm
tr}(\br_\perp)\,e^{-i\bQ \br}\,d^3r,
 \label{F}
 \ee
where
 \be
\Psi_{\rm tr}(\br_\perp)=\int \Phi_{\rm
tr}(\bk_\perp)\,e^{i\bk_\perp\br_\perp}\,\fr{d^2k_\perp}{2\pi} \,.
 \label{Psi-trans}
 \ee
Let us remind that in the standard approach the angular
distribution of scattered particles is determined by the
scattering amplitude $f(\bq)$ proportional to the
Fourier-transform of the potential field $U(r)$ (see
Eq.~\eqref{Born}). The same role for the scattering of the wave
packet plays the quantity $F(\bQ)$, which is the Fourier-transform of
the product of functions $U(r) \,\Psi_{\rm tr}(\br_\perp)$ (see
Eq.~\eqref{F}). From here one can deduce several qualitative
conclusions related to the angular distribution of the scattered
electrons (see Subsection 4.1 below).

\subsection{Averaging over impact parameters}

Let the potential centres be randomly distributed inside a large disk of the radius ${\cal R}\gg a, \sigma_\perp$. In this case the
averaged cross section $d{\bar \sigma}$ is obtained after the
integrating the number of events over all the impact parameters $\bb$
and dividing the result obtained by the total number of particles
in the packet:
 \be
\fr{d{\bar \sigma}}{d\Omega}=\fr{1}{N_e}\int
\fr{d{\nu}}{d\Omega}\,d^2 b,
 \ee
where $d{\nu}/{d\Omega}$ is given by Eq.~\eqref{nuf}.

If the packet axis is shifted in the transverse plane by a distance
$\bb$ from the potential centre , the corresponding wave function
in the momentum representation can be written as
 \be
\Phi_{\rm tr} (\bk_\perp)= a(\bk_\perp) \,e^{-i\bk_\perp\bb},
 \ee
where the function $a(\bk_\perp)$ corresponds to a non-shifted
packet. Therefore, the averaged cross section is proportional to
the integral
 \be
 I_{\rm av}=\int F({\bQ}) F^*({\bQ}) d^2 b,
 \ee
where
 \bea
F(\bQ)&=&\int f(\bQ-\bk_\perp)\, a(\bk_\perp)
\,e^{-i\bk_\perp\bb}\,\fr{d^2k_\perp}{2\pi}\,,
 \\
F^*(\bQ)&=&\int f^*(\bQ-\bk'_\perp)\, a^*(\bk'_\perp)
\,e^{i\bk'_\perp\bb}\,\fr{d^2k'_\perp}{2\pi}\,.
 \eea
After a trivial integration over $\bb$ and $\bk'_\perp$, we obtain
 \be
I_{\rm av}=\int \left|f(\bQ-\bk_\perp)\right |^2 \,\left|\Phi_{\rm
tr} (\bk_\perp) \right |^2 d^2 k_\perp ,
 \label{Iav}
 \ee
and, therefore,
 \be
\fr{d{\bar \sigma}}{d\Omega}=\fr{1}{\cos{\theta_k}}\int
\left|f(\bQ-\bk_\perp)\right |^2 \,
dW(\bk_\perp),\;\;dW(\bk_\perp)=\left|\Phi_{\rm tr} (\bk_\perp)
\right |^2 d^2 k_\perp.
 \label{gav}
 \ee
This expression can be interpreted as averaging of the standard
Born cross section\\ $d\sigma_{\rm st}/d\Omega
=\left|f(\bQ-\bk_\perp)\right |^2$ with the shifted momentum
transfer $\bq\to \bQ-\bk_\perp$ over probability $dW(\bk_\perp)$
to have such a shift in the initial wave packet.

\section{Gaussian wave packet}

In this section we discus the general properties of the Gaussian
wave packets and  derive the basic formulas for models mentioned
in the Introduction. The obtained equations will be used  for
calculation of specific features of these models.

\subsection{General properties}

Let the initial beam be the Gaussian wave packet whose transverse
wave function in the momentum representation is
 \be
\Phi_{\rm tr}(\bk_\perp)= \fr{e^{-(\bk_\perp\sigma_\perp
)^2-i\bk_\perp \bb}}{\sqrt{\pi/(2\sigma_\perp^2)}}\,.
 \ee
Therefore, the dispersion $\Delta k_x =\Delta k_y=
1/(2\sigma_\perp)$ and
 \be
\la \bk_\perp \ra =0,\;\;\la
k_\perp\ra=\varkappa_0=\fr{\sqrt{\pi}}{2\sqrt{2}\sigma_\perp}\approx
\fr{0.63}{\sigma_\perp}\,.
 \ee
Taking into account inequality~\eqref{ineq-p-1} and the relation
$\varkappa_0=p_i \tan{\theta_k}$, we can put below $\theta_k=0$
for the Gaussian wave packet. The coordinate wave function reads
 \be
\Psi_{\rm tr}(\br_\perp, t)= \int \Phi_{\rm tr}(\bk_\perp)
e^{i[\bk_\perp \br_\perp -\bk^2_\perp
t/(2m_e)]}\,\fr{d^2k_\perp}{2\pi}
 \ee
and the transverse density equals
 \be
n_{\rm tr}(\br_\perp,t)=N_e\,|\Psi_{\rm tr}(\br_\perp,t)|^2=
\fr{N_e}{2\pi \sigma_\perp^2(t)}
\,e^{-(\br_\perp-\bb)^2/[2\sigma_\perp^2(t)]}\,.
 \ee
For such a packet $\la \br_\perp \ra = \bb$ and the dispersion
 \be
\Delta x=\Delta
y=\sigma_\perp(t)=\sqrt{\sigma_\perp^2+\left(\fr{t}{2\sigma_\perp
m_e}\right)^2}\,.
 \ee
We use the approximation~\eqref{ineq}, which implies that during
the collision time the transverse dispersion $\sigma_\perp(t)$
almost does not differ from $\sigma_\perp$. In this case, the
function
 \be
\Psi_{\rm tr}(\br_\perp, t)\approx \Psi_{\rm tr}(\br_\perp)=\int
\Phi_{\rm tr} (\bk_\perp)\,e^{i\bk_\perp
\br_\perp}\,\fr{d^2k_\perp}{2\pi}
 \label{psitr}
 \ee
and the quantity $L$ from~\eqref{lum} (at $b=0$) becomes equal to
 \be
L=n_{\rm tr}({\bf 0})=N_e\,\left|\int \Phi_{\rm tr}
(\bk_\perp)\,\fr{d^2k_\perp}{2\pi} \right|^2=\fr{N_e}{2\pi
\sigma_\perp^2}
 \label{lumGa}
 \ee
(compare this expression with~\eqref{Lplane}).

If the Gaussian packet is wide, $\sigma_\perp \gg a$, the behavior
of the function $U(r) \,\Psi_{\rm tr}(\br_\perp)$ in the essential
region of integration in Eq.~\eqref{F} is almost the same as the
one of the potential field $U(r)$. In this case the function
$F(\bQ)$  has almost the same behavior as the standard  Born
amplitude $f(\bq)$. With the decrease of $\sigma_\perp$, behavior of
these two functions becomes more and more different. For example,
the decrease of the function $U(r) \,\Psi_{\rm tr}(\br_\perp)$ with
the growth of $r$ becomes sharper and, therefore, in the
function $F(\bQ)$ role of the larger values of $Q$, compared to the standard Born amplitude $f(\bq)$, increases. As a result,
{\it the angular distribution becomes  wider compared to the standard case}.

Let us show how the standard result follows from Eq.~\eqref{nuf}.
The standard case corres\-ponds to a wide packet which has the
distribution over $\bk_\perp$ concentrated in the narrow region
near $\la k_\perp \ra =\varkappa_0\approx 0$. Therefore, in the
amplitude $f(\bQ-\bk_\perp)$ we can put $\bk_\perp = {\bf 0}$ and
take this amplitude out of the integral over $\bk_\perp$ in the
Eq.~\eqref{nuf}. After that this equation becomes of the form
 \be
\fr{d\nu}{d\Omega}=|f(\bq)|^2 \,N_e\,\left|\int \Phi_{\rm tr}
(\bk_\perp)\,\fr{d^2k_\perp}{2\pi} \right|^2\,,
 \label{res}
 \ee
which coincides with the standard result~\eqref{stannu} for the
number of the scattered particles if we take into account the
relation~\eqref{lumGa}.

Analogously, the averaged cross section for a wide wave packet
coincides with the standard cross section:
 \be
\fr{d{\bar \sigma}}{d\Omega}=\left|f(\bq)\right |^2 \,\int
dW(\bk_\perp)=\left|f(\bq)\right |^2=\fr{d{\sigma_{\rm
st}}}{d\Omega}  \;\;\;\mbox{at}\;\;\;\sigma_\perp \gg a.
 \ee

\subsection{Model 1 -- scattering of the Gaussian wave packet on the Gaussian
potential}

For the potential field of the Gaussian form~\eqref{Gauss}, the
integral $F(\bQ)$ in Eq.~\eqref{nuf} is calculated analytically
 \be
F(\bQ)=B\,\fr{e^{(Q_\perp\,
a)^2/(1+\sigma_\perp^2/a^2)}}{1+a^2/\sigma_\perp^2}\;
\fr{f(\bQ)}{\sqrt{2\pi}\,\sigma_\perp}\,e^{-i\beta}\,,\;\;
Q_\perp=p_i \sin{\theta}\,,
 \label{JGauss}
 \ee
where
 \be
B=e^{-b^2/[4(\sigma_\perp^2+a^2)]}\,,\;\;
\beta=\fr{\bQ_\perp\bb}{1+\sigma_\perp^2/a^2}\,.
 \label{B}
 \ee
The final results for the number of events and for the averaged
cross section have simple analytical forms:
\bea
\label{csGG}
 \fr{d\nu}{d\Omega}&=&B^2\,\fr{e^{2(Q_\perp a)^2
 /(1+\sigma_\perp^2/a^2)}}{(1+a^2/\sigma_\perp^2)^2}\;
\fr{d\nu_{\rm st}}{d\Omega},\;\;\;\fr{d\nu_{\rm
st}}{d\Omega}=L\,\fr{d\sigma_{\rm st}}{d\Omega},\;\;
 L=\fr{N_e}{2\pi\sigma^2_\perp}\,
 \\
 \label{avcsGG}
\fr{d\bar \sigma}{d\Omega}&=&\fr{e^{2(Q_\perp
a)^2/(1+\sigma_\perp^2/a^2)}}{(1+a^2/\sigma_\perp^2)}
\;\fr{d\sigma_{\rm st}}{d\Omega}\,.
 \eea
It is easy to see from this expression that for a wide beam (at
$\sigma_\perp\gg a, b$) we get the standard results.

\subsection{Models 2  -- scattering of the Gaussian wave packet on
the hydrogen atom in the ground state}

The potential field of the hydrogen atom in the ground state is
given by Eq.~\eqref{atom}. In this case the integral $F(\bQ)$ in
Eq.~\eqref{nuf} can be calculated using the substitution
 \be
f(\bQ-\bk_\perp)= f_0\left(\fr{1}{z}+\fr{1}{z^2}\right)=
f_0\,\int_0^\infty\,(1+x)\,e^{-xz}\, dx\,,\;\; z=1+\mbox{$\frac
14$} (\bQ-\bk_\perp)^2\,a^2
 \label{sub}
 \ee
and further simple integration over $\bk_\perp$. As a result, the
differential number of events is expressed via one-fold integral
over  the variable $x$:
 \be
\fr{d\nu}{d\Omega}=L\, f_0^2 \left| \int_0^\infty e^{-xg_0-ig_1 b
\cos{\varphi}-g_2b^2} \,\fr{1+x}{1+x/s}\,dx \right|^2\,,\;\;
 L=\fr{N_e}{2\pi\sigma^2_\perp}\,,\;\;
s=\fr{4\sigma_\perp^2}{a^2}\,,
 \label{Gauss2}
 \ee
 \be
g_0=1+\fr{(Q_\perp a)^2}{4(1+x/s)}+\fr{(Q_z a)^2}{4} \,,\;\;
g_1=\fr{x}{x+s}\,Q_\perp\,,\;\; g_2=\fr{1}{(x+s)a^2}\,.
 \ee
It is  easy to check that Eq.~\eqref{Gauss2} does not change
under the replacement $\varphi \to \pi+\varphi$. It means that the
number of events is symmetric with respect to the angle
$\varphi=\pi$.

In a similar way, the averaged cross section~\eqref{gav} can be
calculated using the substitution
 \be
|f(\bQ-\bk_\perp)|^2=
f^2_0\left(\fr{1}{z^2}+\fr{2}{z^3}+\fr{1}{z^4}\right)=
f^2_0\,\int_0^\infty\,(x+x^2+\mbox{$\frac 16$} x^3)\,e^{-xz}\, dx
 \label{subH}
 \ee
and further simple integration over $\bk_\perp$, which results in
the following expression
 \be
\fr{d\bar\sigma}{d\Omega}={f_0^2}\, \int_0^\infty e^{-xg
}\;\fr{x+x^2+(x^3/6)}{1+x/(2s)}\,dx \,,\;\;g=1+\fr{(Q_\perp
a)^2}{4[1+x/(2s)]}+\fr{(Q_z a)^2}{4}\,.
 \label{Gauss2a}
 \ee

In the limiting case of the wide packet (at $\sigma_\perp\gg a,
b$) we obtain
 \be
g_0=g=1+(qa/2)^2,\;g_1=g_2=0,\; \fr{d\nu}{d\Omega}= L\,f_0^2
\left(\fr{1}{g}+\fr{1}{g^2}\right)^2,\;\;
\fr{d\bar\sigma}{d\Omega}={f_0^2}\left(\fr{1}{g}+\fr{1}{g^2}\right)^2,
 \ee
i.\,e. the standard results.

\subsection{Comparison of models}

In this section we compare scattering of the Gaussian packet
on the Gaussian potential and on the hydrogen atom in the ground
state. All figures in this subsection are calculated at
$p_ia=p_fa=10$ (for the scattering on the hydrogen atom it
corresponds to the energy $\varepsilon_i=1.36$ keV).

\begin{figure}[ht]
\includegraphics[width=0.49\textwidth]{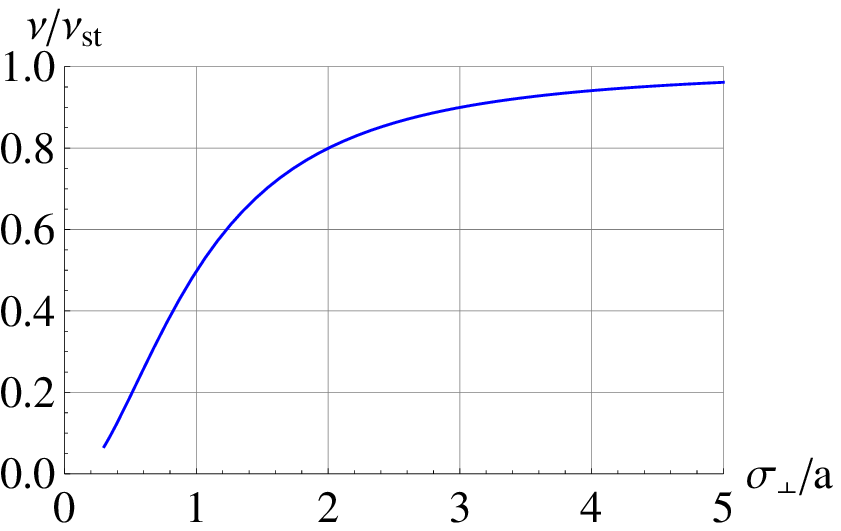}\hfill
\includegraphics[width=0.49\textwidth]{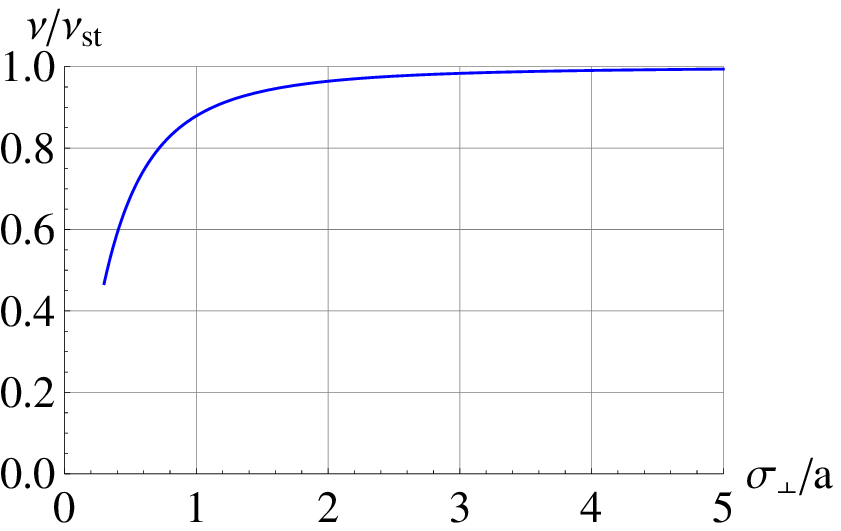}\\
{\caption{\label{Fig1}({\it Left}) Scattering of the Gaussian
packet on the Gaussian potential. Relative number of events vs.
$\sigma_\perp/a$ at $b=0$. ({\it Right}) The same, but for
scattering on the hydrogen atom}}
  \end{figure}

\subsubsection{Central collision (the case ${\bf b}={\bf 0}$)}

Let us consider first the case of the central collision, when the
impact parameter $b=0$. What happens  to the number of events $\nu$
with the decrease of the transverse bunch size $\sigma_\perp$? It is
clearly seen from~Fig.~\ref{Fig1} that the ratio of $\nu$ to the
standard number of events $\nu_{\rm st}$ decreases.
\begin{figure}[h]
    \includegraphics[width=0.49\textwidth]{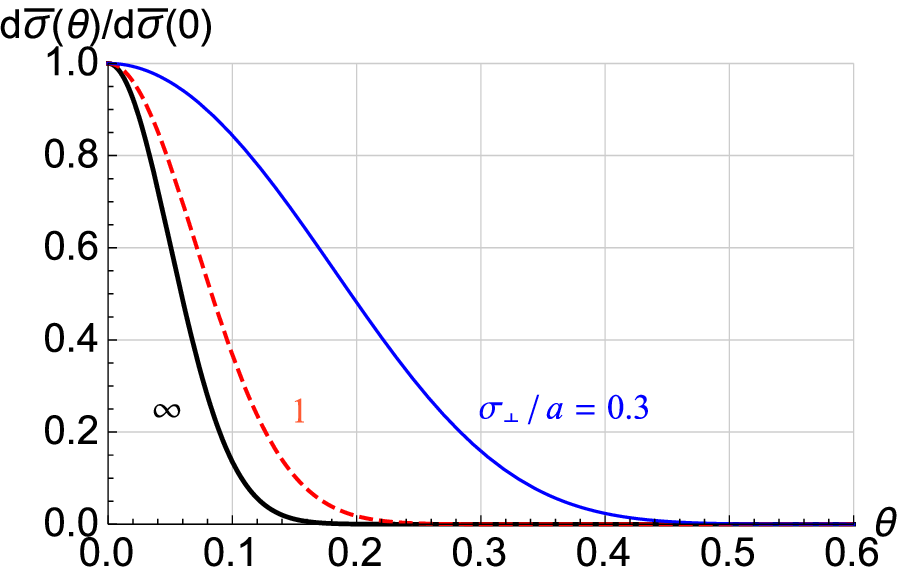}
    \includegraphics[width=0.49\textwidth]{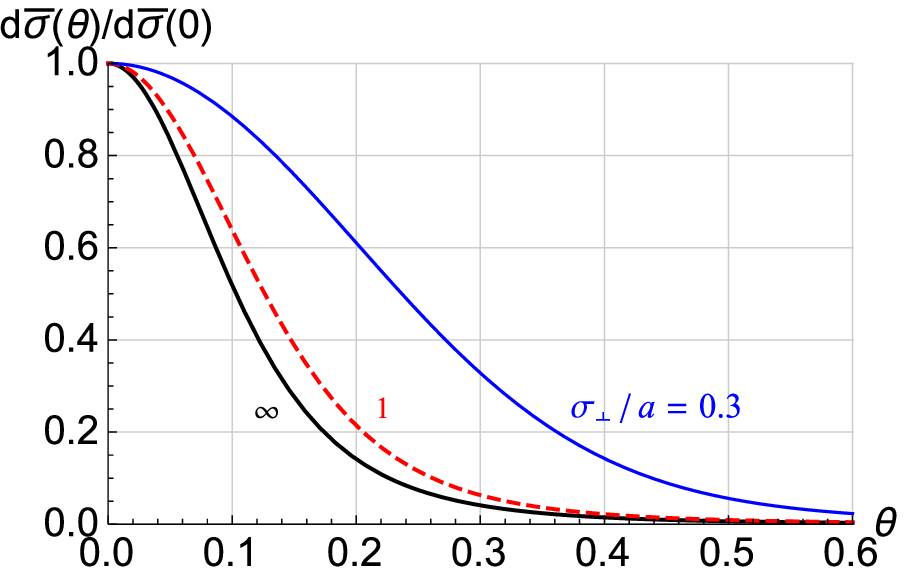}
{\caption{\label{Fig2} ({\it Left}) Scattering of the Gaussian
packet on the Gaussian potential. Relative differential averaged
cross section $[d \bar \sigma(\theta)/ d\Omega] / [d\bar
\sigma(0)/ d\Omega]$ at different values of $\sigma_{\perp} /
a=0.3,\;1,\;\infty$ (from top to below). ({\it Right}) The same,
but for scattering on the hydrogen atom }}
\end{figure}
On the contrary, the angular distribution $d\nu(\theta)/d\Omega$
or $d \bar \sigma(\theta)/ d\Omega$ over the polar angle $\theta$
becomes wider with the decrease of $\sigma_\perp$ (just as we
expected -- see Sect. 4.1). This feature is illustrated by
Fig.~\ref{Fig2}, which presents the relative angular distributions
 \be
\fr{d \bar \sigma(\theta)/ d\Omega}{d \bar\sigma(0)/ d\Omega}
 \label{rel-theta}
 \ee
for $\sigma_{\perp} / a=\infty,\;1,\;0.3$. The physical picture is that  with the decrease of $\sigma_\perp$, the density of the particle current increases in the vicinity of the potential centre and more particles is scattered at larger angles,
but probability of such scattering becomes less (the analytical
expression~\eqref{avcsGG} supports such an interpretation).

\subsubsection{The case ${\bf b}\neq {\bf 0}$}

\begin{figure}[h]
  \centering
\includegraphics[width=0.6\textwidth]{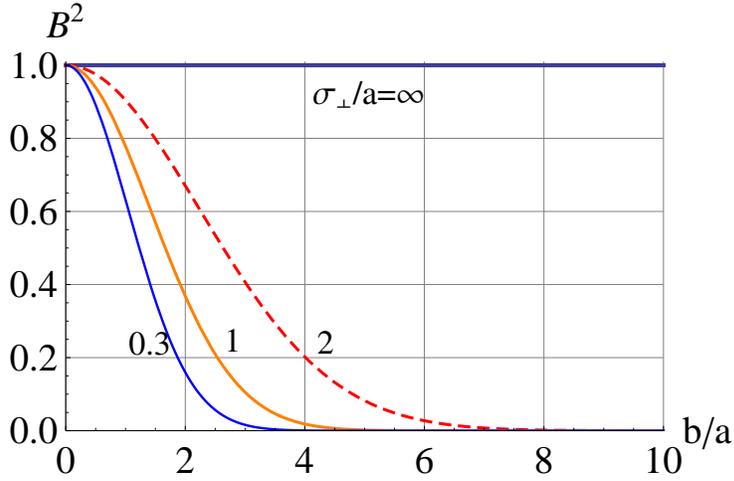}
{\caption{\label{Fig3}Scattering of the Gaussian packet on the
Gaussian potential. The function $B^2$ vs. $b/a$ at
$\sigma_\perp/a= \infty,\, 2,\,1,\,0.3$ (from top to below)}}
 \label{F3}
\end{figure}

The number of events drops quickly with the increase of impact
parameter $b$. It is clearly seen for the model 1 where all the
dependance on $b$ is determined by the function $B^{2}(b)$ (see
Eqs.~\eqref{B} and \eqref{csGG}). This function is presented in
Fig.~\ref{Fig3} for $\sigma_\perp/a= \infty,\, 2,\,1,\,0.3$. It is
seen that this dependence is absent for the standard cross section
(at $\sigma_\perp \gg a, b$). But with the decreasing
$\sigma_\perp$, the observed cross section drops more and more
quickly with the growth of $b$.

From the general point of view we expect that an azimuthal
asymmetry should appear in the angular distribution at $b\neq 0$.
It is quite interesting to note that for the particular model
1 this general expectation is not true! Indeed, though
the quantity $F(\bQ)$ \eqref{JGauss} has the factor $e^{-i\beta}$,
which does depend on the azimuthal angle between vectors
$\bQ_\perp$ and $\bb$, but the number of events, proportional to
$|F(\bQ)|^2$, does not depend on this angle.

\begin{figure}[h]
\centering
\includegraphics[width=0.49\textwidth]{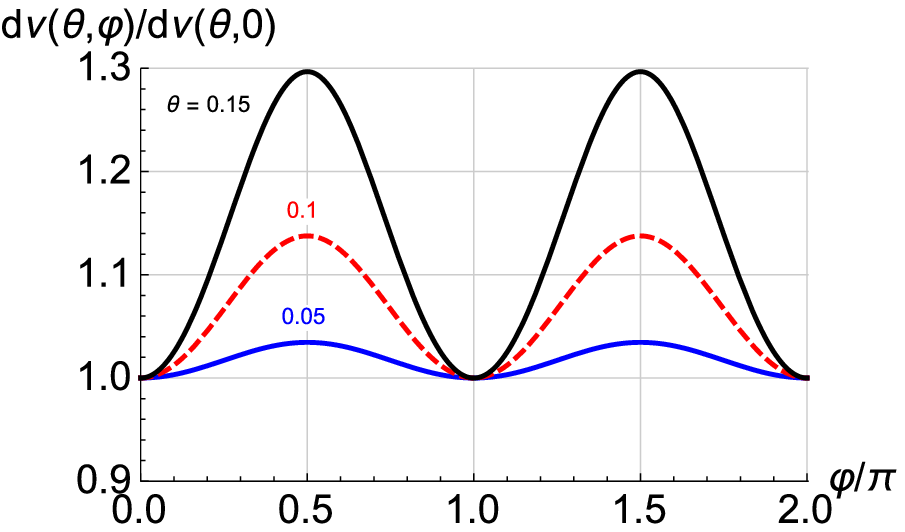}
\includegraphics[width=0.49\textwidth]{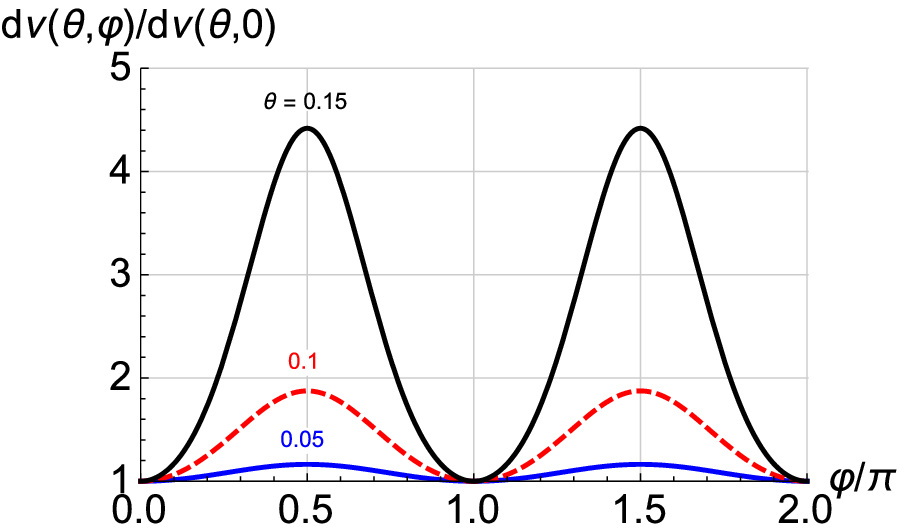}
{\caption{\label{Fig4}({\it Left}) Scattering of the Gaussian
packet with $\sigma_\perp =a$ on the hydrogen atom in the ground
state. Relative differential number of events vs. azimuthal angle
$\varphi$ at $b=2.5a$ and $\theta=0.15,\,0.1,\,0.05$ (from top to
below). ({\it Right}) The same, but for $b=5a$}}
\end{figure}

In contrast to the model 1, the azimuthal asymmetry appears in the
model 2. For this model at $\sigma_\perp=a$ we present in
Fig.~\ref{Fig4} the quantity
 \be
\fr{d\nu(\theta, \varphi)/d\Omega}{d\nu(\theta, 0)/d\Omega}
 \ee
for different values of the polar angle $\theta=0.05,\,
0.1,\,0.15$. The left panel corresponds to the impact parameter
$b=2.5a$, while the right panel --- to the larger impact parameter,
$b=5a$. It is seen that the discussed ratio of cross sections increases considerably at the angles of  $\varphi=\pi/2$ and $\varphi=3\pi/2$
with the growth of the impact parameter.

\section{Summary}

We derived a simple and convenient expression for the number of events, which generalizes the well-known Born approximation for the case when the initial beam
is a well-normalized wave packet, but not a plane wave. Then we considered a couple of simple models corresponding to scattering of the Gaussian wave packet on the Gaussian potential and on the hydrogen atom.

The detailed analysis has been performed of how the total number of events and its angular distributions depend on the limited sizes of the incident beam and on the impact parameters between the potential centre and the packet's axis. In particular, we have found that the angular distributions of the effective cross section broaden with the decrease of the packet's width --- this behaviour is somewhat similar to the pre-wave zone effect in transition radiation, see Ref.\cite{V}. The non-zero impact parameter of the wave packet was shown to lead to azimuthal asymmetry, but, somewhat unexpectedly, this natural effect is absent in the model of the Gaussian potential.

In the next paper we will apply the obtained formulae for scattering of the limited twisted packets on atoms.

\section*{\it Acknowledgements}

We are grateful to A.~Chuvilin, I.~Ginzburg, M.~Entin, I.~Ivanov,
A.~Milstein, and D.~Shapiro for useful discussions. G.L.K and V.G.S. are supported by the Russian Foundation for Basic Research via grant 15-02-05868 and by MES (Russia). D.V.K. is also grateful to the Alexander von Humboldt Foundation for support.

\end{document}